\documentclass[sigconf]{acmart}
\acmConference[ISSTA 2024]{ACM SIGSOFT International Symposium on Software Testing and Analysis}{16-20 September, 2024}{Vienna, Austria}

\copyrightyear{2024}
\acmYear{2024}
\setcopyright{rightsretained}
\acmConference[ISSTA '24]{Proceedings of the 33rd ACM SIGSOFT International Symposium on Software Testing and Analysis}{September 16--20, 2024}{Vienna, Austria}
\acmBooktitle{Proceedings of the 33rd ACM SIGSOFT International Symposium on Software Testing and Analysis (ISSTA '24), September 16--20, 2024, Vienna, Austria}\acmDOI{10.1145/3650212.3680347}
\acmISBN{979-8-4007-0612-7/24/09}

\AtBeginDocument{%
  \providecommand\BibTeX{{%
    Bib\TeX}}}

\usepackage{cleveref}
\usepackage[most]{tcolorbox}
\usepackage{tabularray}
\usepackage{color}
\usepackage{multirow}
\usepackage{enumitem}
\usepackage{amsmath}
\usepackage{amsthm}
\usepackage[skip=2pt, font=small,labelfont=bf]{caption}

\newtheorem{theorem}{Theorem}

\newcommand{\simpy}[0]{\textsc{SimPy }}

\newcommand{\su}[1]{\textcolor{black}{#1}}

\begin{document}

\title{AI Coders Are Among Us: Rethinking Programming Language Grammar Towards Efficient Code Generation}

\author{Zhensu Sun}
\affiliation{%
  \institution{Singapore Management University}
  \country{Singapore}
}
\email{zssun@smu.edu.sg}
\author{Xiaoning Du}
\affiliation{%
  \institution{Monash University}
  \country{Australia}
}
\email{xiaoning.du@monash.edu}
\authornote{Corresponding author.}

\author{Zhou Yang}
\affiliation{%
\institution{Singapore Management University}
\country{Singapore}
}
\email{zyang@smu.edu.sg}
\author{Li Li}
\affiliation{%
  \institution{Beihang University}
  \country{China}
}
\email{lilicoding@ieee.org}
\author{David Lo}
\affiliation{%
\institution{Singapore Management University}
\country{Singapore}
}
\email{davidlo@smu.edu.sg}

\begin{abstract}
Artificial Intelligence (AI) models have emerged as another important audience for programming languages alongside humans and machines, as we enter the era of large language models (LLMs).
LLMs can now perform well in coding competitions and even write programs like developers to solve various tasks, including mathematical problems. 
However, the grammar and layout of current programs are designed to cater the needs of human developers -- with many grammar tokens and formatting tokens being used to make the code easier for humans to read. 
While this is helpful, such a design adds unnecessary computational work for LLMs, as each token they either use or produce consumes computational resources.


To improve inference efficiency and reduce computational costs, we propose the concept of \textit{AI-oriented grammar}.
This aims to represent code in a way that better suits the working mechanism of AI models.
Code written with AI-oriented grammar discards formats and uses a minimum number of tokens to convey code semantics effectively. 
To demonstrate the feasibility of this concept, we explore and implement the first AI-oriented grammar for Python, named Simple Python (\textsc{SimPy}).
\textsc{SimPy} is crafted by revising the original Python grammar through a series of heuristic rules.
Programs written in \textsc{SimPy} maintain identical Abstract Syntax Tree (AST) structures to those in standard Python.
This allows for not only execution via a modified AST parser, but also seamless transformation between programs written in Python and \textsc{SimPy}, enabling human developers and LLMs to use Python and \textsc{SimPy}, respectively, when they need to collaborate.
We also look into methods to help existing LLMs understand and use \textsc{SimPy} effectively.
In the experiments, compared with Python, \textsc{SimPy} enables a reduction in token usage by 13.5\% and 10.4\% for CodeLlama and GPT-4, respectively, when completing the same set of code-related tasks.
Additionally, these models can maintain or even improve their performance when using \textsc{SimPy} instead of Python for these tasks.
With these promising results, we call for further contributions to the development of AI-oriented program grammar within our community. 

\end{abstract}


\maketitle

\section{Introduction}
\label{sec:introduction}

High-level programming languages like Python, Java, and C++ are designed with two types of audiences in mind~\cite{10.1145/3377816.3381720}: machines that compile and execute programs and humans who write, read, and comprehend programs. 
Machines focus on the operational semantics of programs, while humans additionally emphasize programs' readability, which is crucial for understanding source code.
For example, one of the code design principles for Python~\cite{pep20} is that ``readability counts.''
As a result, these languages incorporate many \textit{human-centric} design elements within their grammar. 
For instance, programming languages utilize explicit delimiters to separate code structures, which enhances human readability but may not be essential to convey the program's operational semantics.


Recently, the audiences of programming languages have expanded to include AI models, particularly Large Language Models (LLMs), which can analyze, generate, and execute code.
This is evident form the impressive performance that LLMs achieved in code generation~\cite{codellm_survey}.
For example, AlphaCode2~\cite{alphacode2}, a recently released LLM, reportedly outperforms 85\% of human participants in a programming competition.
Moreover, many LLM-powered assistants, such as ChatGPT~\cite{chatgpt} and Bard~\cite{bard}, are now equipped with code execution environments, which enable them to execute generated code and provide responses based on the results.
Thus, the role of LLMs has evolved from simply being code generators to actively functioning as ``developers'' that use programming to complete a wide range of tasks, such as mathematical computations and file processing.
This shift in paradigm signifies a new era in which AI models emerge as an important group of programming language users.

\begin{figure}
\centering
\includegraphics[width=\linewidth]{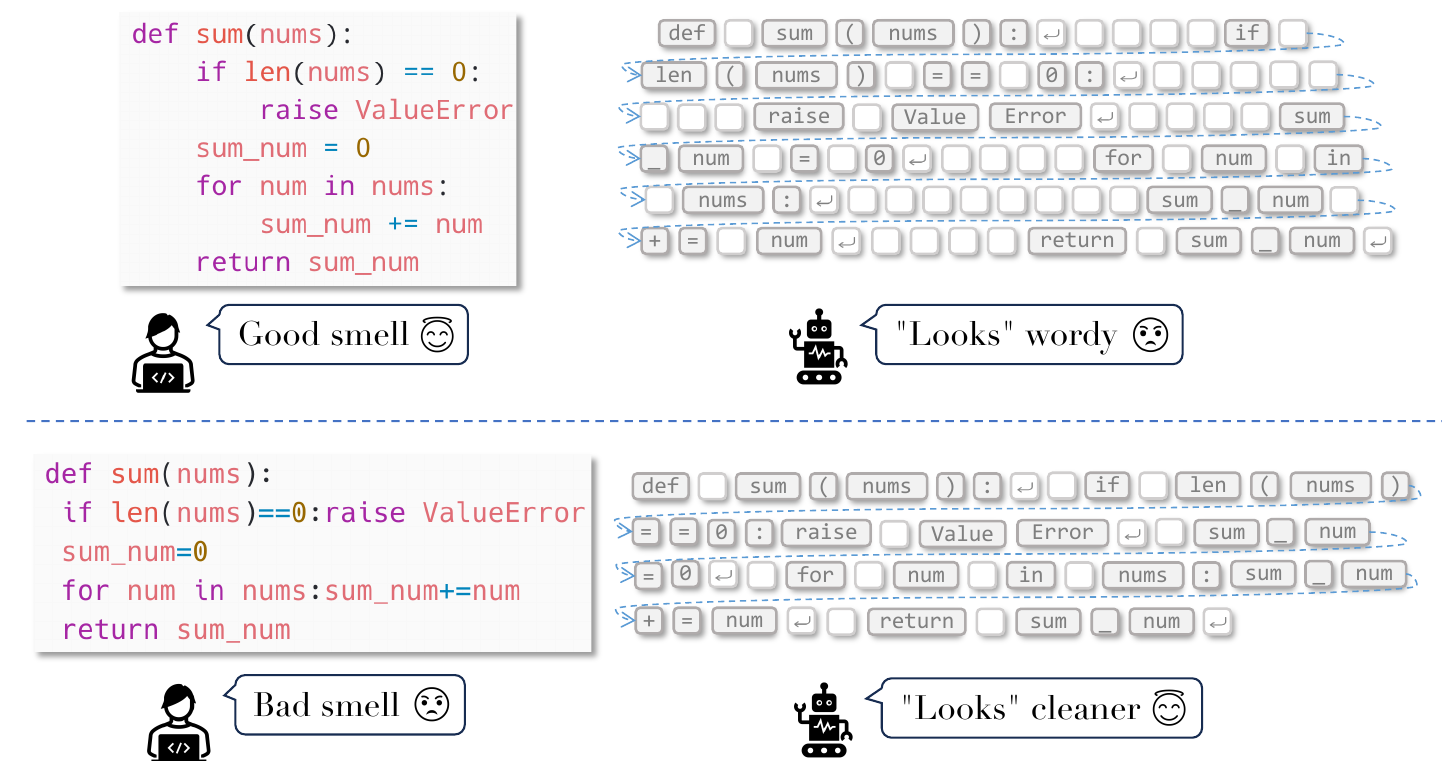}
\vspace{-2mm}
\caption{An illustration of how LLMs and human programmers perceive the source code.}
\vspace{-5mm}
\label{fig:perception_diff}
\end{figure}

While AI models have taken on their new role, the design of code grammar is yet to evolve to accommodate their unique needs.
Elements used to improve the readability of source code could impose an additional computational burden on LLMs when they read and generate programs.
However, the extra readability-enhancing tokens may not be essential for LLMs to perform coding tasks.
Studies have revealed that code models do not capture much information relevant to readability~\cite{Troshin2022ProbingPM}, and readability-enhancing symbols like ``\texttt{:}'' received significantly lower attention compared to other elements such as variable names~\cite{10.1145/3533767.3534390}.
We illustrate how humans and AI models perceive a program in~\Cref{fig:perception_diff}.
When certain elements that improve readability are removed from the code, the program retains its core meaning, but it becomes difficult for humans to understand. 
However, AI models can process the code more efficiently when they are adapted to this new code representation. 
This observation leads us to ask: \textit{What is a suitable grammar for AI models?}
Exploring this question is vital for optimizing the efficiency of LLMs and reducing energy waste in dealing with unnecessary tokens, especially given that the high operational cost of LLMs sets a big challenge for providers to generate profit~\cite{wsj} from them.
As AI models consume and generate source code token-by-token, with one feed-forward process for each token, reducing the tokens in code representation has the potential to reduce the time and energy cost proportionally.

This motivates us to propose the concept of \emph{AI-Oriented Grammar}, a grammar specifically designed for AI models instead of humans.
The core idea is to derive grammar rules that keep the code representations concise (with a minimal/reduced number of tokens to AI models). 
Notably, the code crafted in this grammar can be parsed with its adapted parser and then executed to obtain the same result as the original grammar.
A few challenges are in the way of designing such a new grammar and melting it into AI models.
The AI models are expected to comprehend code written in this grammar and generate code following its rules to better serve the goal of efficiency.
At the same time, human developers, who direct the development, expect to work with grammar that they find friendly and familiar with.

Given these challenges, the realization of this concept remains uncertain. 
To assess the feasibility of AI-oriented grammar, we embarked on an exploratory study.
The study aims to reveal the implications and limitations of integrating AI-oriented grammar into the existing code generation workflow.
Three research questions guide it, each addressing a key challenge.

\vspace*{0.1cm}
\noindent \textbf{RQ1.}
\hangindent=0.8cm \hangafter=1 
\textit{What is the token reduction capacity of AI-oriented grammar in source code?}
\vspace*{0.1cm}

Whether and to what extent an AI-oriented grammar can reduce the tokens remains an open question.
We fill this gap by implementing a proof-of-concept AI-oriented grammar and assessing its performance.
Specifically, we explore a new grammar for Python, named \textsc{SimPy}, by heuristically modifying the Python grammar.
Compared to the standard Python grammar, we prohibit using tokens popularly hired to style the code appearance, e.g., whitespace and newline, and simplify keywords, operators, and delimiters to a more compact form. 
The modifications are designed to be simple, as this is the first attempt to explore such AI-oriented grammar.
We also developed an AST parser for \textsc{SimPy} that can parse its code into the same AST as standard Python code, as well as a converter for seamless code transitions between \textsc{SimPy} and Python code.
A comparative analysis between the \textsc{SimPy} and Python grammars was conducted using tens of tokenizers employed by existing LLMs.
The findings indicate a notable reduction in token usage when employing \textsc{SimPy}, with decreases ranging between 8.6\% and 34.7\%, thus reducing the time and computational cost during inference by a similar level~\cite{Kaplan2020ScalingLF}.
For example, the tokenizer of GPT-4 demonstrates a significantly enhanced efficiency with \textsc{SimPy}, achieving a 10.4\% reduction in token size.

\vspace*{0.1cm}
\noindent \textbf{RQ2.}
\hangindent=0.8cm \hangafter=1 
\textit{How can AI models understand AI-oriented grammar?}
\vspace*{0.1cm}

Prior research demonstrates that AI models can comprehend human-centric grammars of existing programming languages~\cite{codellm_survey}. 
However, how these models can learn AI-oriented grammar remains unexplored.
We thus further experiment with \textsc{SimPy} to find an effective way.
We explored two different training strategies: directly training a model on a \textsc{SimPy}-based code dataset (converted seamlessly from a Python dataset) and fine-tuning a model, originally trained with a Python dataset, on the \textsc{SimPy} dataset.
A control group, where a model is directly trained on the Python code dataset, is also included for comparison. 
The models trained with either strategy should achieve at least equivalent accuracy compared with the control group.
Otherwise, it would be impractical to adopt AI-oriented grammar.
For each training strategy, we experiment with three models: CodeGen-NL, TinyLlama, and Pythia.
The results reveal that models initially trained with Python can adapt effectively to \textsc{SimPy}.
For instance, a CodeGen model, initially trained on Python, attains a 7.32\% Pass@10 on HumanEval; further fine-tuning it on \textsc{SimPy} witnesses an increase of Pass@10 to 9.15\%.

\vspace*{0.1cm}
\noindent \textbf{RQ3.}
\hangindent=0.8cm \hangafter=1 
\textit{How can AI-oriented grammar support real-world scenarios?}
\vspace*{0.1cm}

Given that AI-oriented grammar may compromise human readability, its application is somewhat restricted.
Thus, a remaining challenge for AI-oriented grammar is how it could be used in real-world scenarios, particularly when human-readable source code is necessary.
We first discuss the basic usage scenario of AI-oriented grammar, i.e., scenarios in which the code generated by the AI models is not intended to be displayed to human users. 
AI agents~\cite{Significant_Gravitas_AutoGPT} fall into this category, where the agents solve user-defined problems by generating and executing code in AI-oriented grammar.
The code generated in the process is of little interest to the users.
However, there are still many scenarios in which human developers need to review the code such as collaborative programming with coding assistants.
We thus propose an inference framework for code generation named DualCode.
DualCode utilizes a rule-based converter to convert code between these grammars, ensuring that users interact with human-readable code as usual. At the same time, the model benefits from the efficiency of AI-oriented grammar.
Our experiments confirm that DualCode introduces negligible latency, with the converter of \textsc{SimPy} processing code under 500 tokens in less than 1.0 ms.

The source code of the paper is available at \url{https://github.com/v587su/SimPy}.
The contributions of this paper are summarized as follows:
\begin{itemize}[leftmargin=*]
    \item We propose the concept of AI-oriented grammar and empirically explore its feasibility and potential, paving the way for future improvements in programming language design that prioritize AI efficiency.
    \item We implement the first AI-oriented grammar for Python, named \textsc{SimPy}, which can reduce at least 8.3\% tokens in Python source code.
    \item We propose a novel code generation framework, DualCode, expanding the applicability of AI-oriented grammar beyond AI-only scenarios with negligible additional latency.
\end{itemize}

\section{Motivation}
\label{subsec:ai-oriented-grammar}
In this section, we carefully analyze the human-centered aspects found in the grammar of current programming languages and suggest the idea of an AI-oriented grammar. 
 , we introduce the dataset that later will be used in our study when answering the three RQs.

\subsection{Human-centric Grammar Design}
\label{subsec:human-centric-grammar}

\begin{figure}
    \centering
    \includegraphics[width=\linewidth]{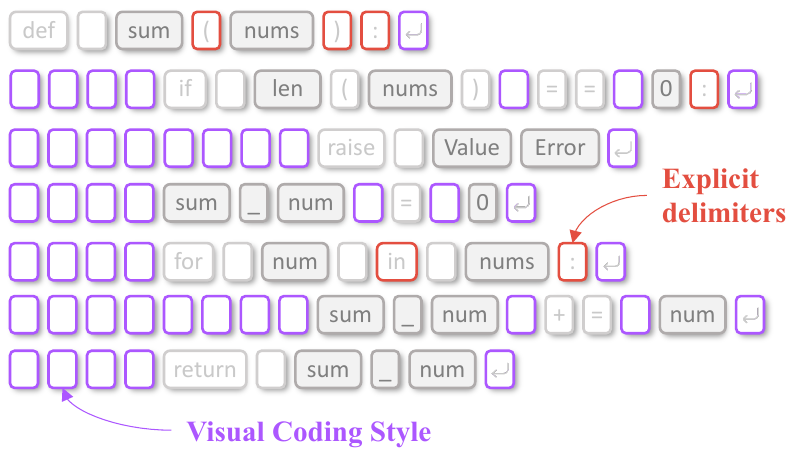}
    \caption{Illustrations of human-centric design elements in Python.}
    \vspace{-2mm}
    \label{fig:design_pattern}
\end{figure}

As discussed in~\Cref{sec:introduction}, modern programming languages are predominantly designed with human-centric grammar.
This design philosophy originates from the longstanding reality that humans were the only developers for decades.
In the current era of LLMs, this human-centric design philosophy has not been significantly challenged.
To better ground this idea, we examine the grammar of widely used programming languages, focusing on lexical and syntactical elements that enhance human readability.
Below, we summarize the identified patterns and provide examples in~\Cref{fig:design_pattern}:

\smallskip \noindent \textbf{Visual Coding Style}
The programming language grammar is deliberately crafted to accommodate diverse coding styles.
Although not mandatory, styles like those recommended in the Python PEP8 guide~\cite{pep8} rely on grammatical support. 
For example, the coding style requires the programs to be written in multiple lines instead of a single extremely long line, easing human code review on screens.
This necessitates several lexical elements: line breaks to separate lines, indents to visualize code blocks, and line continuation symbols for splitting long lines.
\Cref{fig:design_pattern} demonstrates these aspects, with line breaks and indents highlighted in purple.
Similarly, the coding style suggests surrounding each binary operator with a single white space on either side.
Therefore, lexical grammar must accommodate such stylistic elements, even if they may not contribute to the core semantics in parsing.

\smallskip \noindent \textbf{Intuitive Notations}
The human-centric syntax of programming languages is designed to be intuitively understandable to humans.
Common operators like ``+'' for addition and ``='' for assignment are chosen for their familiarity, and derivations like the augmented assignment operator ``+='' maintain this intuitive connection. 
Although potentially more concise symbols could replace these (e.g., using a brand-new symbol ``\$'' for ``+=''), they are still deliberately designed to maintain human readability.
Similarly, for structural clarity, programming languages often employ explicit delimiters, such as symbols or keywords, to define code structures despite these delimiters not being essential for parsing. 
For instance, Python’s compound statements, such as the if statement and for statement, use a colon to demarcate the header from the body.
While a parser might deduce these components from line breaks alone, the colon acts as a visual aid, as illustrated in~\Cref{fig:design_pattern}, where colons are highlighted in red. 
This emphasis on intuitive notation and explicit delimiters, although not essential for parsing, significantly aids human comprehension.

\subsection{AI-Oriented Grammar}
Grammar is a rule set that defines how the source code should describe the programming language's semantics in aspects of lexis and syntax, using notations such as symbols and keywords.
The primary function of the notations in the grammar is two-fold: to define a program’s structure for machine execution and to enhance visual comprehension for human readability.
Given that AI models do not require assistance in visual comprehension, the focus of AI-oriented grammar is solely on structural definition.
We thus consider a notation unnecessary for AI models if it does not contribute to accurate parsing by the parser.
AI-oriented grammar is designed with indispensable notations.

In the design process of a programming language, semantics are defined first, followed by the development of a grammar to represent them. 
Therefore, employing AI-oriented grammar does not alter the fundamental semantics of the programming language.
Technically, a programming language grammar and its AI-oriented design share the AST (abstract syntax tree) definition.
Given that an AST uniquely encodes the semantics of a code, it facilitates an equivalent transformation between implementations of the semantics with either grammar.

\vspace{-1mm}
\subsection{Python Code Dataset for Our Study}
\label{sec:dataset}
As a newly proposed concept, we are still unclear whether AI-oriented grammar can be realized and what scenarios it can be applied to.
To address these uncertainties and explore the potential of AI-oriented grammar, we conduct an empirical study guided by three critical research questions, respectively introduced in~\Cref{sec:ai-oriented-grammar},~\Cref{sec:rq2},~\Cref{sec:framework}.
Our study is centered around Python, the main programming language of the execution environment for LLMs like GPT-4 and Bard to address programming-required tasks.
We utilize the Python subset of starcoderdata~\cite{Li2023StarCoderMT}, a filtered variant of The Stack dataset~\cite{Kocetkov2022TheStack}, a comprehensive collection of over 20 million code files sourced from open-source GitHub repositories. 
We keep the code files from the repositories with over 100 stars, resulting in 623,887 code files.
The dataset is partitioned into training and validation sets in a 95:5 ratio.
We do not create a separate testing set, as we plan to evaluate the model's performance using other established evaluation datasets.
The code snippets in the evaluation datasets are excluded from the training dataset.

\vspace{-1mm}
\section{Token Reduction (RQ1)}
\label{sec:ai-oriented-grammar}
\su{In this section, we present an instance of AI-oriented grammar to answer \textbf{RQ1}: \textit{What is the token reduction capacity of AI-oriented grammar in source code?}}
We propose an AI-oriented grammar for Python as a proof-of-concept (\Cref{subsec:simpy}) and then proceed to evaluate the extent of token reduction achievable with this grammar (\Cref{subsec:rq1_exp}).

\begin{figure}
    \centering
    \includegraphics[width=\linewidth]{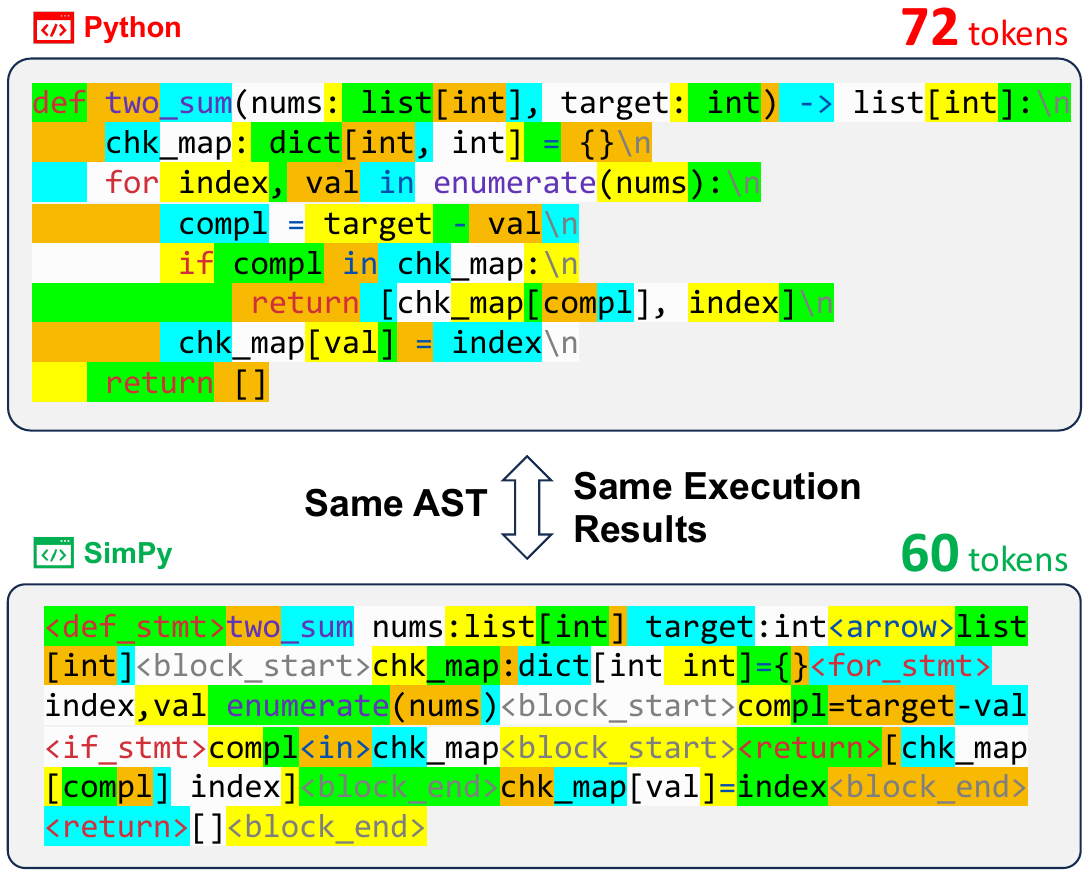}
    \vspace{-3mm}
    \caption{A comparison between Python and \textsc{SimPy} source code, tokenized by GPT-4's tokenizer. Continuous characters with the same background color represent the same token. Notably, there are no line breaks in the \textsc{SimPy} example and we add these line breaks in the figure for our human readers.}
    \vspace{-5mm}
    \label{fig:simpy}
\end{figure}

\begin{table*}
    \caption{Comparison of grammar specifications for Python and \textsc{SimPy}, using the official Python grammar notation (\cite{python-grammar}). 
    Each terminal notation affected by \textsc{SimPy} is annotated in blue.
    The table also includes the count of such terminal notations: ``N'' represents the number of lines, ``n'' signifies the count of repetitive elements, and ``?'' indicates that the number of terminals is conditional.}
    \scalebox{0.88}{\begin{tblr}{
  row{1} = {c},
  colsep = 4pt,
  cell{2}{1} = {r=2}{c},
  cell{2}{2} = {c},
  cell{2}{3} = {fg=blue},
  cell{2}{4} = {c},
  cell{3}{2} = {c},
  cell{3}{3} = {fg=blue},
  cell{3}{4} = {c},
  cell{4}{1} = {r=2}{c},
  cell{4}{2} = {c},
  cell{4}{4} = {c},
  cell{5}{2} = {c},
  cell{5}{4} = {c},
  cell{6}{1} = {r=2}{c},
  cell{6}{2} = {c},
  cell{6}{4} = {c},
  cell{7}{2} = {c},
  cell{7}{4} = {c},
  cell{8}{1} = {r=2}{c},
  cell{8}{2} = {c},
  cell{8}{4} = {c},
  cell{9}{2} = {c},
  cell{9}{4} = {c},
  cell{10}{1} = {r=2}{c},
  cell{10}{2} = {c},
  cell{10}{4} = {c},
  cell{11}{2} = {c},
  cell{11}{4} = {c},
  cell{12}{1} = {r=2}{c},
  cell{12}{2} = {c},
  cell{12}{4} = {c},
  cell{13}{2} = {c},
  cell{13}{4} = {c},
  cell{14}{1} = {r=2}{c},
  cell{14}{2} = {c},
  cell{14}{4} = {c},
  cell{15}{2} = {c},
  cell{15}{4} = {c},
  cell{16}{1} = {r=2}{c},
  cell{16}{2} = {c},
  cell{16}{4} = {c},
  cell{17}{2} = {c},
  cell{17}{4} = {c},
  cell{18}{1} = {r=2}{c},
  cell{18}{2} = {c},
  cell{18}{4} = {c},
  cell{19}{2} = {c},
  cell{19}{4} = {c},
  cell{20}{1} = {r=2}{c},
  cell{20}{2} = {c},
  cell{20}{4} = {c},
  cell{21}{2} = {c},
  cell{21}{4} = {c},
  vlines,
  hline{1-2,4,6,8,10,12,14,16,18,20,22} = {-}{},
}
\textbf{Name} & \textbf{Grammar} & \textbf{Specification} & \textbf{\#Terminal}\\
block & Python & \textcolor{blue}{NEWLINE INDENT} \textcolor{black}{statements} \textcolor{blue}{DEDENT}~ & 3\\
 & \textsc{SimPy} & \textcolor{blue}{`<block\_start>'} \textcolor{black}{statements} \textcolor{blue}{`<block\_end>'} & 2\\
function\_def & Python & \textcolor{blue}{`def'} NAME [type\_params] \textcolor{blue}{`{}('} [params] \textcolor{blue}{`{})'} [\textcolor{blue}{`->'} expression ] \textcolor{blue}{`{}:'} [func\_type\_comment] block & 4+1?\\
 & \textsc{SimPy} & \textcolor{blue}{`<def\_stmt>'} NAME [type\_params] [params] [\textcolor{blue}{`<arrow>'} expression ] [func\_type\_comment] block & 1+1?\\
class\_def & Python & ~\textcolor{blue}{`{}class'} NAME [\textcolor{blue}{`('} [arguments] \textcolor{blue}{`{})'} ] \textcolor{blue}{`{}:'} block~ & 2+2?\\
 & \textsc{SimPy} & ~\textcolor{blue}{`{}<class\_stmt>'} NAME [\textcolor{blue}{`('} [arguments] \textcolor{blue}{`{})'} ]~ block~ & 1+2?\\
if\_stmt & Python & ~\textcolor{blue}{`{}if'} named\_expression \textcolor{blue}{`{}:'} block elif\_stmt~ & 2\\
 & \textsc{SimPy} & ~\textcolor{blue}{`{}<if\_stmt>'} named\_expression block elif\_stmt~ & 1\\
for\_stmt & Python & ~\textcolor{blue}{`{}for'} star\_targets \textcolor{blue}{`{}in'} \textasciitilde{} star\_expressions \textcolor{blue}{`{}:'} [TYPE\_COMMENT] block [else\_block]~ & 3\\
 & \textsc{SimPy} & ~\textcolor{blue}{`{}<for\_stmt>'} star\_targets \textasciitilde{} star\_expressions [TYPE\_COMMENT] block [else\_block]~ & 1\\
with\_stmt & Python & ~\textcolor{blue}{`{}with'} \textcolor{blue}{`{},'}.with\_item+ \textcolor{blue}{`{}:'} [TYPE\_COMMENT] block~ & 2+n\\
 & \textsc{SimPy} & ~\textcolor{blue}{`{}<with\_stmt>'} \textcolor{blue}{`{} '}.with\_item+ [TYPE\_COMMENT] block~ & 1\\
try\_stmt & Python & ~\textcolor{blue}{`{}try'} \textcolor{blue}{`{}:'} block except\_block+ [else\_block] [finally\_block]~ & 2\\
 & \textsc{SimPy} & ~\textcolor{blue}{`{}<try\_stmt>'} block except\_block+ [else\_block] [finally\_block]~ & 1\\
while\_stmt & Python & ~\textcolor{blue}{`{}while'} named\_expression \textcolor{blue}{`{}:'} block [else\_block]~ & 2\\
 & \textsc{SimPy} & ~\textcolor{blue}{`{}<while\_stmt>'} named\_expression block [else\_block]~ & 1\\
import\_from & Python & ~\textcolor{blue}{`{}from'} (\textcolor{blue}{`.'} \textbar{} \textcolor{blue}{`{}...'})* dotted\_name \textcolor{blue}{`{}import'} import\_from\_targets & 2+n?\\
 & \textsc{SimPy} & ~\textcolor{blue}{`{}<from\_import\_stmt>'} (\textcolor{blue}{`.'} \textbar{} \textcolor{blue}{`{}...'})* dotted\_name import\_from\_targets & 1+n?\\
simple\_stmts & Python & ~\textcolor{blue}{`{};'}.simple\_stmt+ [\textcolor{blue}{`;'}] \textcolor{blue}{NEWLINE} & n+1+1?\\
 & \textsc{SimPy} & ~[\textcolor{blue}{`<line\_sep>'}].simple\_stmt+ [\textcolor{blue}{`<line\_sep>'}] & n?+1?\\
\end{tblr}}
    \label{tab:grammar}
    \vspace{-4mm}
\end{table*}

\vspace{-1mm}
\subsection{An AI-oriented grammar for Python}
\label{subsec:simpy}
We introduce how \textsc{SimPy}, a simplified grammar for Python, is designed by applying the philosophy of AI-oriented grammar in~\Cref{sssection:design}.
We discuss how \textsc{Simpy} is validated against ambiguity in~\cref{sssection:unambiguity} and prove the semantic equivalence in~\cref{sssection:semantic-equ}.
Alongside \textsc{SimPy}, we develop a toolkit including a parser to interpret \textsc{SimPy} source code into Python's AST, and a converter for seamless code translation between \textsc{SimPy} and Python.

\subsubsection{Design}
\label{sssection:design}
A straightforward method to reduce the number of tokens in source code is to remove redundant delimiters. For example, consecutive newline symbols or whitespace (neither indent nor dedent) can be reduced to a single instance without affecting the code's semantics. These tokens can be optimized in programs written in any programming language. Inspired by the example in Introduction, we explore slight modifications to an existing grammar to reduce even more tokens.

As the first attempt at AI-oriented grammar design, we limit the changes to terminals in grammar; the semantics of production rules are kept intact.
Terminals are notations\footnote{Terminals are tokenized by the compiler lexer. To distinguish the tokenization performed by the lexer and that performed by the tokenizers of LLMs, we use the term ``notation'' to refer to the tokens produced by the lexer.} that cannot be expanded by any production rule.
Typical terminals include keywords (e.g., ``true''), symbols (e.g., ``>='', ``('', and ``,''), literals (e.g., user-defined constants), and identifiers (e.g., variable names).
Limiting the changes to terminals additionally helps maintain semantic equivalence between Python and \simpy, minimizes potential ambiguity in \simpy, and keeps implementation efforts at an acceptable level.

In the following, we use Python 3.12 as the base grammar and describe the modifications made to obtain \simpy. 
\textsc{SimPy} is not guaranteed to be optimal in terms of model-processing efficiency but is sufficient to serve as a proof-of-concept demonstration for AI-oriented grammar.
For a quick comparison, an illustration of \textsc{SimPy} and Python code is shown in~\Cref{fig:simpy}, where both code snippets share the same AST but the \simpy representation consists of fewer tokens (measured by the GPT-4 tokenizer).
In~\Cref{tab:grammar}, we compare the grammar specifications of key productions before and after these modifications and count the tokens affected in the modification. 
Limited by the space, we introduce the two major categories of modifications here; the complete grammar specification is available in our artifact.

\noindent \textbf{Replace terminal notations with tokens.}
This type of modification is based on two observations: 1) the whitespace around certain terminals, such as ``true'', is mandatory to ensure recognizability, and 2) some terminals, like ``+='', might be tokenized into multiple tokens. If we can find a way to eliminate the need of surrounding whitespace for these terminals and ensure that they are parsed as a single token, more tokens can be removed when squeezing out redundant delimiters.
Based on whether the terminals are made of a fixed sequence of characters, they can be divided into two groups, namely constant terminals and user-defined terminals.
The instances of user-defined terminals are not available to the grammar definition.
Here, we propose to replace the constant terminals by distinct token placeholders.
For example, ``true'' and ``>='' will be replaced by ``<true>'' and ``<ge>'', respectively.
In this way, whitespace around these placeholders is no longer needed.
During the tokenization by LLMs, each placeholder will be mapped directly to an entry in the vocabulary of the tokenizer, being treated as a single token.
Note that some single-character symbols, like ``.'', ``('', 
and ``='', are not replaced in consideration of the common optimization provided by the popular tokenizers -- these symbols are combined into their next token thus contributing no extra token.
For example, ``(nums'' in~\Cref{fig:design_pattern} is treated as a single token by GPT-4.



\noindent \textbf{Simplify notations in the grammar context.}
Some terminal notations can be further simplified in the context of specific production rules.
During the design, we review every production rule and determine if any notations can be removed, merged with or replaced by others.
Here, we mainly focus on delimiter symbols, keywords, and other terminals produced by the lexer (e.g., ``NEWLINE'').
Delimiters separate different program structures, many of which are used to aid humans in reading.
In a given grammar context, some delimiters can be removed without affecting the parsing.
Taking the ``function\_def'' rule as an example, ``('' and ``)'', surrounding the parameters, and ``:'', setting aside the function header and body, are discarded.
As a result, four terminals plus one terminal in the optional structure in the original Python grammar are simplified into one terminal plus one in the optional structure.

In another example, the ``block'' statement in Python hires ``NEWLINE'', ``INDENT'', and ``DEDENT'' to indicate the start and end of a block.
Here, ``NEWLINE'' and ``INDENT'' are merged and replaced by the new token placeholder ``<block\_start>'', while ``DEDENT'' is replaced by ``<block\_end>'' according to the previous design rule.
Three affected terminals in Python grammar are reduced into two.
Another design we would like to highlight is that the``NEWLINE'' terminal, indicating the line breaks, are replaced with an new optional token placeholder ``<line\_sep>'' in simple\_stmts.
It permits the omission of ``<line\_sep>'' when the subsequent a token implying the start of a new line, such as ``<def\_stmt>'' (the replacement of ``def'') for function definitions or ``<class\_stmt>'' (the replacement of ``class'') for class definitions.



Finally, we leverage the design of existing tokenizers to replace some tokens with those that are combined into their following tokens by the tokenizers.
For example, the whitespace character is combined into its following token by GPT-4, as illustrated by `` two'' and `` int'' in~\Cref{fig:design_pattern}.
For some mandatory explicit delimiters, if they can be replaced by a whitespace, the number of tokens can also decrease.
For instance, we replace ``,'' with a whitespace in the with\_stmt.
However, such replacement may cause conflicts, as whitespace is widely used for separation in for many structures.
One example is the list structure in Python, which uses ``,'' to separate different elements, e.g., ``{[}`1', `2'{]}''.
When we replace the ``,'' with a whitespace, the parser generator raises a conflict regarding the concatenation of strings separated by whitespace, such as ```hello' `world'''.
To resolve this conflict, we introduce the string concatenation with a new token placeholder, ``<concat>'', as a separator for string concatenation. This approach considers that the list structure is used more frequently in code. Representing more frequent components with fewer tokens is a widely used strategy in content compression. Designing a grammar that considers the frequency of different terminals, structures, and grammar rules could be a very interesting direction for future work.

\vspace{-2mm}

\subsubsection{Unambiguity of \simpy}
\label{sssection:unambiguity}

\su{
To determine whether a grammar has ambiguity is theoretically undecidable~\cite{Ginsburg1966AmbiguityIC}.
In practice, parser generator tools are commonly hired to check for ambiguities in grammar, including those of popular programming languages~\cite{Grune2007ParsingT}.
A parser generator can find a wide range of ambiguities in the grammar, such as conflicts that arise when the parser has two possible actions at one step. 
Practically, this is almost the best way to check the ambiguity of \textsc{SimPy}.
We have successfully generated parsers for \textsc{SimPy} using the GLR (generalized left-to-right rightmost derivation parser) parsing algorithm~\cite{Lang1974DeterministicTF} from tree-sitter~\cite{tree-sitter}, where no ambiguity is detected.}

\su{
Next, we provide an analytical discussion about why our transformations are unlikely to introduce ambiguity to the grammar.
First of all, the transformations are only made to terminal notations, which act as keywords, symbols, or delimiters.
Changes made to keywords and symbols are guaranteed to represent its unique semantics, while changes made to delimiters should not affect the recognition of the construct, as well as its precedent and subsequent constructs.}

\noindent \textbf{Case I: }
\textit{New notations are added or introduced as replacements.}

\su{
Importantly, different notations are not replaced with the same new notations.
To this end, the new notations do not interfere with production rules for which the transformation is not applicable.
Given that they are semantically equivalent notations as the original one, the parsing of the affected production rules remains the same.
For example, replacing the \textit{`NEWLINE INDENT'} in the production rule of block (see~\Cref{tab:grammar}) with \textit{`<blcok\_start>'} conveys the same semantics that a block is about to start.}

\noindent \textbf{Case II:}
\textit{Existing notations are removed.}

\su{
Arbitrary removal of notations may introduce ambiguity to the grammar. We carefully design a few heuristics when removing notations so they are unlikely to cause problems.
\begin{itemize}[leftmargin=*]
\item \textbf{Remove notations with redundant semantics as their adjacent notations.}
For example, \textit{‘:’} in many statements indicates the end of the previous construct and the start of a new construct, e.g., in \textit{‘if’ named\_expression ‘:’ block elif\_stmt}.
However, the block construct initiates with its starting symbol, making the construct itself distinguishable from any previous construct. Hence, removing \textit{‘:’} is safe for this case.
\item \textbf{Remove delimiters used to scope a construct when the scope of its precedent and subsequent constructs is clear.}
For example, the \textit{‘(’} and \textit{‘)’} for parameters are actually unnecessary in \textit{function\_def\_raw := ‘def' NAME [type\_params] ‘(’ [params] ‘)’ [‘->’ expression ] ‘:’ [func\_type\_comment] block}. \textit{NAME} is an atomic token, thus will not interfere with the beginning of parameters when \textit{type\_params} is absent.
\textit{type\_params} are surrounded by \textit{‘[’} and \textit{‘]’}, making their presence not an issue for recognizing params. Hence, \textit{‘(’} can be safely removed.
Now, looking at the subsequent constructs, \textit{[‘->’ expression ]}, \textit{‘:'}, \textit{[func\_type\_comment]}, or \textit{block} possesses a unique indicator of their beginning. Hence, \textit{‘)’} can be safely removed as well. 
Another example is the ‘\textit{import}’ keyword in \textit{import\_from := ‘from’ (‘.’ | ‘...’)* dotted\_name ‘import’ import\_from\_targets}.
Since \textit{dotted\_name} is a must and contains no white spaces, hence the white space between \textit{dotted\_name} and \textit{import\_from\_targets} can perfectly separate these two constructs.
Removing ‘\textit{import}’ is also fine.
\end{itemize}}
\vspace{-2mm}

\subsubsection{Semantic equivalence between SimPy and Python}
\label{sssection:semantic-equ}

\su{\simpy is designed as a simplified grammar of Python, which means a program written in Python can be equivalently and deterministically transformed to its counterpart in \simpy, and vice versa.
In other words, Python and \simpy are semantically equivalent.
We prove this statement in~\Cref{thm:target}.}

Formally, we define a grammar $G$ and a grammar $G'$. 
$G'$ is obtained via a transformation $T$ to the production rules in $G$.
Given a production rule, $T$ is restricted to adding, replacing, or removing a terminal notation or a sequence of terminal notations.
The transformation between Python and \simpy is an instance complying with this restriction.
For example, \textit{$T$(NEWLINE INDENT statements DEDENT ) = `<block\_start>’ \(T\)(statements) `<block\_end>’}.

\vspace{-2mm}
\begin{theorem}
     \su{Python and $\simpy$ are semantically equivalent.}
     \label{thm:target}
\end{theorem}
\vspace{-2mm}

\begin{proof}
Two programs are semantically equivalent if they share the same AST. 
We assume the Python grammar can be transformed into the \simpy grammar via \(T\).



We give the proof by structural induction on \(p\).

\textbf{Base case:} \(p\) is an atomic program construct. 
According to the design of \(T\), for keywords and symbols used in atomic program construct, a subset of them are transformed to a new token, e.g., \(true\) is transformed to <\(true\)>.
Since the mapping is unique and deterministic, a keyword and its mapped token share identical semantics and will be abstracted to the same AST node.

\textbf{Inductive case:} This can be proved by analyzing each compound language construct that is affected by the transformation.
Due to the space limitation, we only show one proof for the \textit{if\_stmt} construct.
Assuming \textit{\(p\) = `if' named\_expression `:' block elif\_stmt}, we have \textit{\(p'\) = `<if\_stmt>' T(named\_expression) T(block) T(elif\_stmt)}.
Both \(p\) and \(p'\) will be translated into an AST rooted with a node representing the \(if\) statement, and with three children representing the \textit{named expression}, \textit{function body}, and \textit{else statement}.
By the induction hypothesis, \(T(named\_expression)\), \(T(block)\), and\(T(elif\_stmt)\) share the same AST with \(named\_expression\), \(block\), and \(elif\_stmt\), respectively.
Hence, \(p\) and \(p'\) share the same AST.

By proving all other constructs, we can prove that for any program in Python, its counterpart in \simpy is semantically equivalent to it.
Similarly, we can prove that for any program in \simpy, its counterpart in Python is semantically equivalent to it as well.
Thus, the theorem is proved.
\end{proof}



\subsubsection{Implementation}
Based on the grammar specifications of \textsc{SimPy}, we develop a toolkit for it, including an AST parser for \textsc{SimPy} code and a converter for seamless translation between \textsc{SimPy} and Python source codes. 
The parser is built upon tree-sitter~\cite{tree-sitter}, a popular parser generator tool. 
We first describe the grammar specification of \textsc{SimPy} in the configuration file of the tree-sitter and then generate the parser.
With the help of the GLR algorithm from the tree-sitter, we ensure \textsc{SimPy} resolves all the conflicts and no ambiguity exists.
The generated parser can parse the \textsc{SimPy} source code into the AST of Python.
Based on this parser, we further implement a converter, where specific conversion rules are established for each node of the AST.
\su{
From a pragmatic point of view, we test our implemented toolkits by conducting round-trip transformations, where Python source code is first converted into \textsc{SimPy} code and subsequently retranslated back to Python. 
Our first tests on the Python dataset revealed that, ignoring all whitespace, the textual content of the code remains unchanged after the transformation. 
In addition, we assess its soundness through execution results.
We perform the round-trip transformation to the ground-truth code snippets of HumanEval and run the test cases on both the transformed and the original code.
The execution results of all the transformed code and the original code are exactly the same, which also indicates the soundness of our implementation.}


\subsection{Experiments of RQ1}
\label{subsec:rq1_exp}
In this section, we detail the tokenizers employed in our experiments and describe the experimental results.

\subsubsection{Tokenizers}
Our experiments encompass a broad spectrum of tokenizers from various LLMs.
The main difference between them is the training corpus, leading to different token vocabularies.

\noindent \textbf{GPT-2}~\cite{Radford2019LanguageMA}, \textbf{Codex}~\cite{Chen2021EvaluatingLL}, \textbf{GPT-3.5}~\cite{gpt3.5}, \textbf{GPT-4}~\cite{OpenAI2023GPT4TR}: These tokenizers, released by OpenAI, are trained on a mixed corpus, including both natural language and programming language, with GPT-4 being the latest version offering state-of-the-art performance in various language tasks.

\noindent \textbf{CodeLlama}~\cite{Rozire2023CodeLO}, \textbf{WizardCoder}~\cite{Luo2023WizardCoderEC}, \textbf{DeepSeek-Coder}~\cite{deepseek-coder}: These tokenizers are derived from the tokenizer of Llama 2~\cite{Touvron2023Llama2O} which is also trained on the mixed corpus.

\noindent \textbf{SantaCoder}~\cite{allal2023santacoder}, \textbf{StarCoder}~\cite{Li2023StarCoderMT}, \textbf{Replit-code}~\cite{repilt}: These tokenizers are specialized for code, having been trained exclusively on programming language datasets, and are thus more adept at handling source code.

\noindent \textbf{CodeGen}~\cite{Nijkamp2022CodeGenAO}, \textbf{CodeT5}~\cite{Wang2021CodeT5IU}, \textbf{CodeT5+}~\cite{Wang2023CodeT5OC}: These tokenizers are extended based on the vocabulary of GPT2 with additional tokens representing repeating tokens of tabs and white spaces.

\begin{table}
    \centering
    \setlength{\tabcolsep}{3pt}
    \caption{Percentage of token reduction achieved with \textsc{SimPy}. The ``Code'' and ``Web'' in the ``Vocab Source'' column represent the sources for constructing the tokenizer's vocabulary: code repositories and internet data, respectively.}
    \scalebox{0.88}{\begin{tblr}{
  cells = {c},
  colsep = 4.5pt,
  cell{1}{1} = {r=2}{},
  cell{1}{2} = {r=2}{},
  cell{1}{3} = {r=2}{},
  cell{1}{4} = {c=3}{},
  cell{2}{5} = {c=2}{},
  vline{1-5,7} = {-}{},
  hline{1,3,17} = {-}{},
  hline{2} = {4-6}{},
}
\textbf{Tokenizer} & {\textbf{Vocab}\\\textbf{Source}} & {\textbf{Vocab}\\\textbf{Size}} & \textbf{Tokens} &  & \\
 &  &  & \textbf{Python} & \textbf{\textsc{SimPy}} & \\
CodeBert & Code & 50k & 1.33B & 0.87B & 34.7\%$\downarrow $\\
GPT2 & Web & 50k & 1.33B & 0.87B & 34.7\%$ \downarrow $\\
CodeLlama & Web & 32k & 0.97B & 0.84B & 13.5\%$ \downarrow $\\
WizardCoder & Web & 32k & 0.97B & 0.84B & 13.5\%$ \downarrow $\\
DeepSeek-Coder & Web & 32k & 0.97B & 0.84B & 12.9\%$ \downarrow $\\
CodeGen & Web & 51k & 0.93B & 0.82B & 12.6\%$ \downarrow $\\
CodeT5+ & Web & 51k & 0.93B & 0.82B & 12.6\%$ \downarrow $\\
Codex & Web & 51k & 0.93B & 0.82B & 12.6\%$ \downarrow $\\
CodeT5 & Code & 32k & 0.91B & 0.78B & 13.8\%$ \downarrow $\\
StarCoder & Code & 49k & 0.83B & 0.76B & 8.6\%$ \downarrow $\\
SantaCoder & Code & 49k & 0.83B & 0.76B & 8.8\%$ \downarrow $\\
Replit-code & Code & 33k & 0.82B & 0.75B & 8.6\%$ \downarrow $\\
GPT-3.5 & Web & 100k & 0.71B & 0.63B & 10.4\%$ \downarrow $\\
GPT-4 & Web & 100k & 0.71B & 0.63B & 10.4\%$ \downarrow $
\end{tblr}

}
    \vspace{-2mm}
    \label{tab:token_count}
\end{table}

\subsubsection{Results}
To answer RQ1, we conducted an evaluation involving the representation of code files from our Python dataset in both its original grammar and in \textsc{SimPy}, followed by the tokenization using the same tokenizer for each representation. 
We created the \textsc{SimPy} dataset by converting the Python dataset with our converter. 
In tokenizing the \textsc{SimPy} code, we modify the tokenizers to include tokens of \textsc{SimPy} in their vocabularies. 
In total, 14 tokenizers from popular LLMs are evaluated in our experiments, where each tokenizer's vocabulary source and size are also documented to offer a comprehensive view of \textsc{SimPy}'s performance across different models.
By examining the variation in token numbers, we evaluated \textsc{SimPy}'s effectiveness in reducing token size, thus showcasing the potential benefits of AI-oriented syntax. 

As revealed in \cref{tab:token_count}, \textsc{SimPy} can reduce the number of tokens by 8.6\% to 34.7\%, depending on the tokenizers.
The GPT-4 and GPT-3.5 tokenizers, which are already the most efficient in representing Python source code, show a further reduction of 10.4\% in token count with \textsc{SimPy}. 
For tokenizers trained on code corpora, such as Replit-code and StarCoder, \textsc{SimPy} achieved a token reduction ranging from 8.6\% to 13.8\%. 
Tokenizers trained on web-based corpora like CodeGen and CodeT5 also exhibited significant reductions, between 12.6\% and 13.5\%.
The most pronounced impact of \textsc{SimPy} is observed with the least efficient tokenizers, CodeBert and GPT-2, where a remarkable 34.7\% reduction in token count was achieved.
These promising results highlight \textsc{SimPy}'s potential to reduce token count for source code representation.
As estimated by OpenAI~\cite{Kaplan2020ScalingLF}, the Floating-point operations (FLOPS) required for generating each token during inference can be regarded as being only relevant to the model size when the context size is fixed.
Therefore, a reduction in token count can be directly translated to a decrease in FLOPS at a similar level, resulting in faster inference speeds given the fixed computing speed of the device.

\begin{tcolorbox}[size=title]
    {\textbf{Answer to RQ1:}}
    AI-oriented grammar, exemplified using \textsc{SimPy}, effectively reduces the number of tokens required for source code representation, with models like GPT-4 benefiting from a 10.4\% reduction. Correspondingly, it leads to a speed up and a computing saving during inference at a similar level.
\end{tcolorbox}

\section{Model Training (RQ2)}
\label{sec:rq2}
\su{In this section, we aim to answer \textbf{RQ2}: \textit{How can AI models understand AI-oriented grammar}?}
We experimentally investigate whether AI models can retain their accuracy when trained with AI-oriented grammar.
We describe our training strategies in \Cref{subsec:adaption} and assess their effectiveness on two language models in \Cref{subsec:rq2_exp}.

\subsection{Training Strategies}
\label{subsec:adaption}
Training AI models with AI-oriented grammar is a pivotal step in enabling the model to deal effectively with source code in this new format.
Despite the efficiency gains demonstrated by \textsc{SimPy}, such training should not compromise the model's accuracy. 
To explore the feasibility of such training, we experiment with two different strategies.
Next, we introduce the strategies in the experiment, from tokenizer refining to model training.

\noindent \textbf{Tokenizer Refining}
\textsc{SimPy} introduces 78 new tokens for the tokenizers to recognize.
For example, the ``def'' keyword of the original Python grammar is replaced by a token ``<def\_stmt>''.
Given the existing association between the pre-trained model and its tokenizer, completely retraining the tokenizer on \textsc{SimPy} code to optimize token distribution is impractical.
Instead, we opt for a more feasible approach: expanding the tokenizer's vocabulary to include these new tokens.
Correspondingly, this modification requires resizing the embedding matrix ({}[vocab size * embedding size]) and the output layer ({}[hidden state size * vocab size]) to fit the expended vocab size.
This expansion introduces a few new parameters, mainly in the output layer, around 78 * hidden\_size parameters.
For instance, modifying a CodeGen~\cite{Nijkamp2022CodeGenAO} model with a hidden state size of 2048 introduces around 160 thousand new parameters, a negligible increase (less than 0.01\%) in the total parameter count.
Moreover, the resizing will randomly initialize both the embedding vector for each new token and the weight of the output layer, which will be updated during the model training.

\noindent \textbf{Model Training}
Our study explores two basic training strategies: 1) directly training a model on the \textsc{SimPy} code dataset, referred to as \textbf{\textsc{SimPy}}, and 2) sequentially training a model first on the Python dataset and then on the \textsc{SimPy} code dataset, referred to as \textbf{Python$\rightarrow$\textsc{SimPy}}. 
If such basic strategies work, further improvement in efficiently adapting AI-oriented grammar is completely feasible.
Moreover, we construct a control group: directly training a model on the Python code dataset, denoted as \textbf{Python}.
The performance of the two strategies should match or surpass the model from the control group; otherwise, they are not practical.
To control the variable, all training sessions across the two strategies and the control group are conducted under identical conditions, including the training environment, initial model, and training hyper-parameters. 
Notably, the \textsc{SimPy} dataset is converted from the Python dataset, ensuring no external data is involved.
Moreover, for the Python+\textsc{SimPy} setting, we vary the proportion of the \textsc{SimPy} dataset used, i.e., 10\%, 20\%, 50\%, and 100\%, to assess the required volume of data for effective fine-tuning.
\begin{table}
    \centering
    \setlength{\tabcolsep}{2pt}
    \caption{The Pass@1 and Pass@10 of LLMs on Python and \textsc{SimPy} datasets under varied settings. Python and \textsc{SimPy} denote models trained exclusively on respective datasets. Python$\rightarrow$\textsc{SimPy} refers to sequential training on both datasets, with the parenthetical numbers indicating the \textsc{SimPy} dataset's proportion involved in the training.}
    \scalebox{0.88}{\begin{tblr}{
  cells = {c},
  cell{2}{1} = {r=6}{},
  cell{8}{1} = {r=6}{},
  cell{14}{1} = {r=6}{},
  vlines,
  hline{1-2,8,14,20} = {-}{},
  hline{3-4,9-10,15-16} = {2-4}{},
}
Model & Training Strategy & Pass@1 & Pass@10\\
CodeGen-NL & ~Python & 4.51\% & 7.32\%\\
 & ~100\% \textsc{SimPy} & 2.93\% & 5.49\%\\
 & ~Python $\rightarrow$ 10\% \textsc{SimPy} & 3.11\% & 3.66\%\\
 & ~Python $\rightarrow$ 20\% \textsc{SimPy} & 3.66\% & 4.27\%\\
 & ~Python $\rightarrow$ 50\% \textsc{SimPy} & 3.96\% & 6.71\%\\
 & ~Python $\rightarrow$ 100\% \textsc{SimPy} & \textbf{4.82\%} & \textbf{9.15\%}\\
TinyLlama & ~Python & 10.00\% & 13.41\%\\
 & ~100\% \textsc{SimPy} & 5.91\% & 9.76\%\\
 & ~Python $\rightarrow$ 10\% \textsc{SimPy} & 2.07\% & 3.66\%\\
 & ~Python $\rightarrow$ 20\% \textsc{SimPy} & 3.23\% & 5.49\%\\
 & ~Python $\rightarrow$ 50\% \textsc{SimPy} & 5.73\% & 11.59\%\\
 & ~Python $\rightarrow$ 100\% \textsc{SimPy} & \textbf{10.12\%} & \textbf{14.02\%}\\
Pythia & ~Python & 5.79\% & 9.76\%\\
 & ~100\% \textsc{SimPy} & \textbf{7.01\%} & 9.15\%\\
 & ~Python $\rightarrow$ 10\% \textsc{SimPy} & 1.89\% & 2.44\%\\
 & ~Python $\rightarrow$ 20\% \textsc{SimPy} & 3.11\% & 4.27\%\\
 & ~Python $\rightarrow$ 50\% \textsc{SimPy} & 4.21\% & 7.32\%\\
 & ~Python $\rightarrow$ 100\% \textsc{SimPy} & 5.67\% & \textbf{10.00\%}
\end{tblr}}
    \vspace{-4mm}
    \label{tab:performance}
\end{table}

\vspace{-1mm}
\subsection{Experiments of RQ2}
\label{subsec:rq2_exp}
We first present the experimental setup for RQ2, including the models used, evaluation metrics, and implementation details. 
Then, we report the experimental results and answer the research questions.

\subsubsection{Models}
\su{
We adopt three widely used models in our research community, namely CodeGen-NL, TinyLlama, and Pythia, whose parameter sizes range between 350M and 1.1B.
}
All these models serve as the initial pre-trained model for our experiments.
Though these are not the latest state-of-the-art models, they suffice to validate the feasibility of learning AI-oriented grammar like \textsc{SimPy}.
We will further discuss the impact of this decision in~\Cref{sec:threat}.

\begin{figure*}
    \centering
    
    \includegraphics[width=0.6\linewidth]{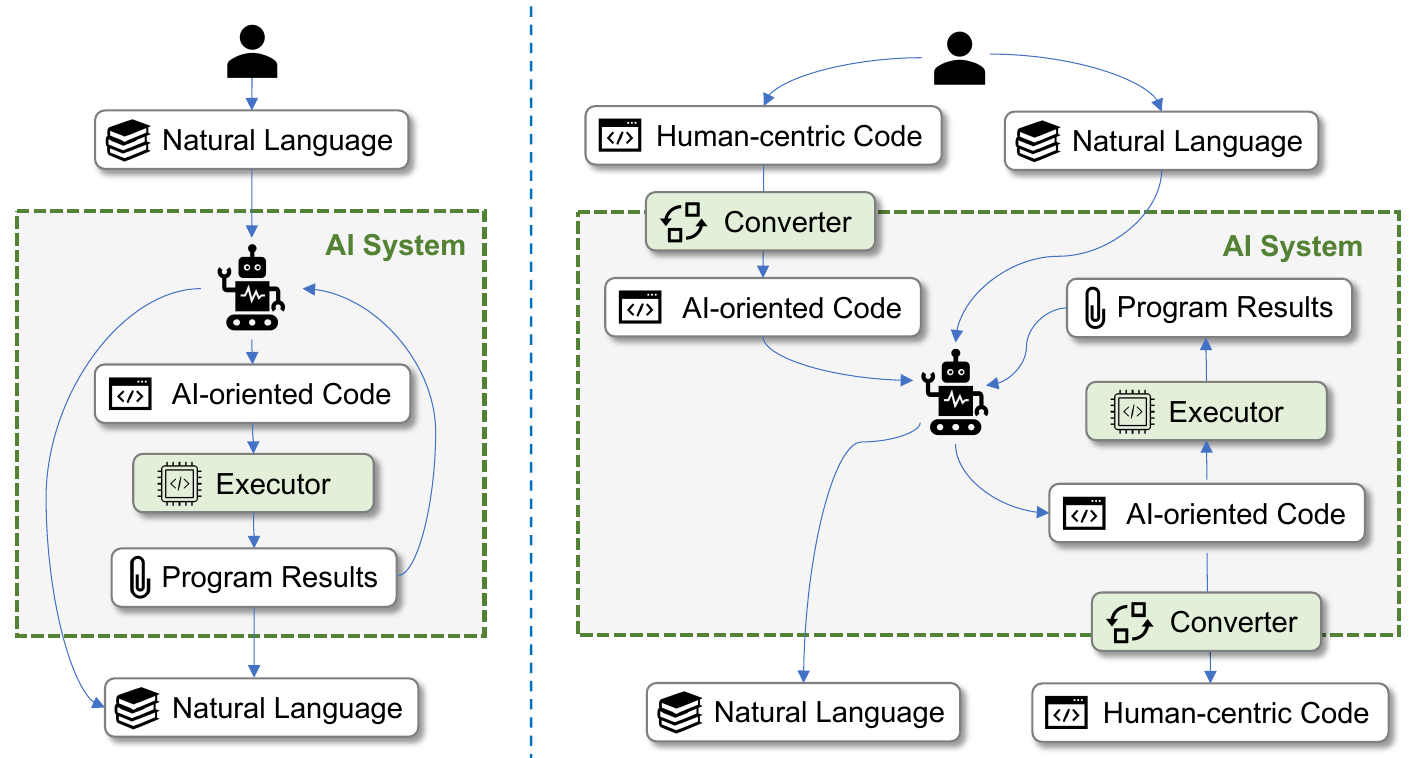}
    
    \caption{LEFT: the workflow of the basic usage scenarios of AI-oriented grammar. RIGHT: the workflow of the extended usage scenarios of AI-oriented grammar under DualCode, where the code executor of the AI system in the figure is not necessary.}
    \vspace{-3mm}
    \label{fig:worflow}
\end{figure*}

\noindent \textbf{CodeGen-NL}: CodeGen, proposed by Salesfore~\cite{Nijkamp2022CodeGenAO}, is an open-sourced language model designed for code generation.
It undergoes a multi-phase training process on different datasets, where the model is first trained with natural language datasets and then code datasets.
Our experiments utilize its natural language version (CodeGen-350M-nl), produced after the initial phase of its training process, as the foundation model to conduct our experiments.

\noindent \textbf{TinyLlama}: \su{TinyLlama~\cite{zhang2024tinyllama} is a compact 1.1B language model pre-trained on around 3 trillion tokens, building on the architecture and tokenizer of Llama 2~\cite{Touvron2023Llama2O}.
It shows competitive performance compared to existing open-source language models of similar sizes.}

\noindent \textbf{Pythia}: \su{Pythia~\cite{biderman2023pythia} is a suite of LLMs, which is expected to be used as the baseline for research studies and thus is designed close to currently accepted common practices.
Considering the capacity of our computing resources, we use its 1B version.
}

\subsubsection{Evaluation Metrics}
We evaluate the model's performance on the code generation task with the Pass@$k$ metric on HumanEval.
\su{To compute Pass@$k$, $k$ code samples are generated for each problem, and a problem is considered solved if any of the $k$ samples pass the unit tests. We report the fraction of problems being successfully solved.}
The HumanEval dataset~\cite{Chen2021EvaluatingLL} comprises 164 programming problems, each with a function signature, a docstring, and multiple test cases. 
Given the function signature and docstring, the model is required to generate the code, which is then tested by executing the test cases.
Notably, the function signatures are written using Python's original grammar.
When evaluating the model adapted to \textsc{SimPy}, we convert the function signature into \textsc{SimPy} using the code converter.
\su{Similarly, the model-generated \textsc{SimPy} code is subsequently converted into Python to run test cases since the existing testing framework is implemented for Python source code.}

\subsubsection{Implementation Details}
In our experiments, we use the Huggingface Transformers library~\cite{wolf-etal-2020-transformers} with Pytorch to implement the models.
\su{The experiments of CodeGen-NL are performed on a machine with 48 vCPUs, 512GB RAM, and four RTX A5000 GPUs {(24GB RAM)}, while the other two models are trained on a machine with 28	vCPUs, 200GB RAM, and two RTX A6000 GPUs {(48GB RAM)}.}
The hyper-parameters of the training are set referring to CodeGen's hyper-parameters: 8 batch size, 1.8e-4 learning rate, 0.1 weight decay, and 512 context length.
During the inference for evaluation, we set the temperature to 0.2 and the top-p to 0.95.

\subsubsection{Results}

Following the settings of the two strategies (\textsc{SimPy} and Python$\rightarrow$\textsc{SimPy}) and the control group (Python), we train the CodeGen-NL, \su{TinyLlama, and Pythia} models, respectively.
Finally, for each of our initial models, we have six variations: one each for Python and \textsc{SimPy}, and four models for Python$\rightarrow$\textsc{SimPy} incorporating 10\%, 20\%, 50\%, and 100\% of the \textsc{SimPy} dataset. 
The performance of these models is evaluated through Pass@1 and Pass@10 metrics on the HumanEval dataset.

We report the results in~\Cref{tab:performance}.
Notably, the models trained with \textsc{SimPy} lag behind the Python baseline in terms of accuracy.
For example, the Pass@1 and Pass@10 of CodeGen (\textsc{SimPy}) are respectively 2.93\% and 5.49\%, lower than the ones of CodeGen (Python), which are 4.51\% and 7.32\%.
This could be attributed to \textsc{SimPy}'s limited expressiveness, constraining the models from leveraging knowledge acquired from natural language datasets during pre-training.
Consequently, direct training with AI-oriented grammar appears to be an impractical approach.

However, the sequential training strategy, starting with Python and then incorporating \textsc{SimPy}, yields comparable or even superior accuracy to the control group. 
\su{
Specifically, CodeGen-NL, TinyLlama, and Pythia models trained with Python$\rightarrow$100\%\textsc{SimPy} achieve Pass@10 scores of 9.15\%, 14.02\%, and 10.00\%, respectively, outperforming the control group's 7.32\%, 13.41\%, and 9.76\%.
}
This suggests a successful training with \textsc{SimPy}, demonstrating the feasibility of AI models learning AI-oriented grammar.
\su{Interestingly, we observe that the Pythia model, when trained exclusively with 100\% \textsc{SimPy}, surpasses the Python baseline on Pass@1.
This highlights the possibility of learning \textsc{SimPy} without relying on the sequential training strategy.}
By varying the proportion of the \textsc{SimPy} dataset in the Python$\rightarrow$\textsc{SimPy} setting, we found that a substantial dataset is still required by the fine-tuning with \textsc{SimPy}.
\su{
For instance, TinyLlama (Python$\rightarrow$50\%\textsc{SimPy}) scored 5.73\% in Pass@1 and 11.59\% in Pass@10, still trailing behind the TinyLlama (Python) scores.
}
We will further discuss this finding in~\Cref{sec:discussion}.

\begin{tcolorbox}[size=title]
{\textbf{Answer to RQ2:}} AI models, when initially trained with the original grammar and then the AI-oriented grammar, can successfully learn the AI-oriented grammar while retaining their accuracy.
For instance, the CodeGen model, originally trained with Python and achieving a 7.32\% Pass@10, improved to a 9.15\% Pass@10 after the additional training with \textsc{SimPy}.
\end{tcolorbox}
\section{Usage Scenario (RQ3)}
\label{sec:framework}
\su{In this section, we address \textbf{RQ3}: \textit{How can AI-oriented grammar support real-world scenarios?}}
We first demonstrate the basic application scenario of AI-oriented grammar, and subsequently introduce a novel inference framework designed to broaden the applicability of AI-oriented grammar, followed by an evaluation of the framework's additional latency.

\vspace{-1mm}
\subsection{Basic usage scenario}
The source code, when written in AI-oriented grammar, becomes challenging for human interpretation and is therefore not intended for human display.
Consequently, the application of AI-oriented grammar is limited to scenarios where human users do not have access to the generated code.
A typical scenario is the AI agents, such as AutoGPT~\cite{Significant_Gravitas_AutoGPT} and LangChain~\cite{Chase_LangChain_2022}, for regular users rather than developers.
For instance, an AI agent tasked with data collection from a website would generate the required crawler script, execute it to gather data, and present the outcomes to the user.
End users generally care more about the results than understanding the underlying script since they lack programming knowledge.
Therefore, even without additional enhancement, models trained with AI-oriented grammar can be effectively utilized in real-world scenarios.
We demonstrate this scenario on the left of~\Cref{fig:worflow}.
\su{
In this scenario, an AI-oriented code generated by the model can be executed in two ways: 1) being translated into human-centric code and then executed by its executor; 2) directly being executed by a specific executor for the AI-oriented grammar.
Notably, implementing an executor specifically for AI-oriented grammar demands only lightweight engineering efforts as the AI-oriented grammar and its original grammar differ only at the syntax level. 
Thus, the second method offers a more efficient solution.
}
\vspace{-1mm}

\subsection{Extended usage scenario}
Despite the effectiveness of AI-oriented grammar in certain contexts, many code generation scenarios still require the involvement of humans, where human-readable code is required.
To fill this gap, we propose an inference framework for code generation named DualCode.
DualCode enables human users to interact with code in human-centric grammar, while the model still leverages the efficiency of AI-oriented grammar during the inference process.
The fundamental concept of DualCode is to convert the code between AI-oriented grammar and the original grammar of the same programming language.
To achieve this goal, a rule-based code converter should be employed to convert source code into AI-oriented grammar for model comprehension and, inversely for user readability.
Such a converter is feasible since both the AI-oriented grammar and original grammar describe the same AST.
The identical AST allows the code written in the two grammars to be equivalently converted into each other based on the grammar rules.

We illustrate the workflow of DualCode on the right of~\Cref{fig:worflow}.
It employs two ``gates'': an input converter and an output converter.
The input converter translates code written in human-centric grammar into AI-oriented grammar for model processing.
Similarly, the output converter reverts AI-generated code into human-readable code for user comprehension.
Notably, this environment is only for the code, where other inputs, such as natural language, are unaffected.
\su{DualCode is a not complicated framework, enabling the lightweight integration of AI-oriented grammar into existing workflows of AI systems.
Though being straightforward, it is proposed and investigated for the first time, 
bridging the gap between efficient AI-oriented code generation and human readability.
}

\vspace{-1mm}
\subsection{Experiments of RQ3}
\begin{table}
    \centering
    \setlength{\tabcolsep}{3pt}
    \caption{Comparison of average conversion times between Python and \textsc{SimPy}, and the processing speed of the StarCoder tokenizer, based on Huggingface Tokenizers.}
    \scalebox{0.88}{\begin{tblr}{
  cells = {c},
  colsep = 5pt,
  cell{1}{1} = {r=2}{},
  cell{1}{2} = {c=2}{},
  cell{1}{4} = {c=2}{},
  vline{1-3,5} = {1}{},
  vline{4,6} = {2}{},
  vline{1-2,4,6} = {1-7}{},
  hline{1,3,8} = {-}{},
  hline{2} = {2-5}{},
}
\textbf{Token num} & \textbf{Huggingface} &  & \textbf{Converter} & \\
 & \textbf{Encode} & \textbf{Decode} & \textbf{To \textsc{SimPy}} & \textbf{To Python}\\
{[}0, 100) & 0.2ms & 0.1ms & 0.2ms & 0.2ms\\
{[}100, 500) & 0.7ms & 0.6ms & 0.9ms & 0.8ms\\
{[}500, 2000) & 2.4ms & 2.2ms & 3.4ms & 3.1ms\\
{[}2000, 5000) & 6.7ms & 6.4ms & 12.2ms & 10.8ms\\
{[}5000, +$\infty$) & 23.0ms & 23.7ms & 75.4ms & 57.4ms
\end{tblr}}
    \vspace{-4mm}
    \label{tab:converter}
\end{table}

Given that the DualCode converter adds extra latency to the inference process, a significant concern arises: excessive latency could render the system impractical for real-world applications.
To address the concern, we conduct experiments focusing on the converter's performance. 
Specifically, we measure the time taken to convert Python code files into \textsc{SimPy} and then back to Python using the converter.
As a reference, we evaluate the processing speed of the StarCoder tokenizer, which is based on the widely acknowledged Huggingface Tokenizers library~\cite{Moi_HuggingFace_s_Tokenizers_2023}.
For this experiment, we categorized Python code files into five distinct groups, based on their token counts, as follows: {[}0, 100), {[}100, 500), {[}500, 2000), {[}2000, 5000), and {[}5000, +$\infty$).
\su{These token counts are determined using the StarCoder tokenizer~\cite{Li2023StarCoderMT} on the Python code.}
We calculate the average processing time for each group.

The findings, presented in Table~\ref{tab:converter}, indicate that the converter's speed is comparable to that of Huggingface Tokenizers.
For code files with fewer than 100 tokens, the converter's processing time for each conversion is a mere 0.2 ms, only 0.1 ms slower than the Huggingface Tokenizers. 
For files containing 100 to 500 tokens, the conversion is completed within 1.0 ms.
This is not a significant concern, given that over 95\% of the dataset's code files (sourced from real-world repositories) are within the 5000-token range.
Therefore, we deduce that the latency induced by the converter is acceptably minimal in most practical scenarios.

\begin{tcolorbox}[size=title]
    {\textbf{Answer to RQ3:}} Beyond the basic scenarios where human interaction is not required, the application of AI-oriented grammar can be substantially extended by incorporating the DualCode framework. DualCode enables humans to continue using human-centric grammar while AI models leverage the efficiency of AI-oriented grammar. Notably, it imposes negligible latency (under 1 ms for code up to 500 tokens).
\end{tcolorbox}

\section{Related Work}
\noindent \textbf{Program Simplification}
Program simplification has emerged as a valuable approach to enhance the efficiency of code models~\cite{Huang2023ProgramTV,Rabin2022SyntaxguidedPR,Rabin2021UnderstandingNC, Shi2023StructuralsemanticsGP, Bui2019AutoFocusIA, Yang2024RobustnessSP}.
This approach typically involves the elimination of less critical code tokens to streamline model processing.
For example, DietCode~\cite{Zhang2022DietCI} removes the code tokens that receive the fewest attention weights by CodeBert.
Sivand~\cite{Rabin2022SyntaxguidedPR} and P2IM~\cite{Zheng2020ProbingMS} simplify the input code according to the outputs of a supplementary model.
While these methods considerably boost efficiency, they unavoidably compromise accuracy due to the removal of certain code elements.
In contrast, models with AI-oriented grammar, though perhaps less efficient, are able to preserve or even improve accuracy. 
Most importantly, existing simplification techniques are irreversible, limiting their application to code understanding tasks like summarization and retrieval, rather than code generation.
Conversely, code in AI-oriented grammar can be effortlessly reverted to its original form, thus suitable for various code-related tasks.

\noindent \textbf{Tokenization of Source Code}
Modern LLMs usually preprocess textual datasets using an open-vocabulary tokenization method, Byte-Pair Encoding {(BPE)}~\cite{Sennrich2015NeuralMT}.
BPE tokenizes text into subwords based on their frequency in the text corpus, offering a balance between the granularity of tokens and vocabulary breadth.
Karampatsis et al.~\cite{Karampatsis2020BigC} first identify the effectiveness of BPE on source code.
CodeT5 reveals that BPE trained on source code corpus can reduce over 30\% of tokens for code generation, compared with the one trained on natural language corpus.
Subsequently, all major LLMs for code generation, such as CodeBERT~\cite{Feng2020CodeBERTAP}, CodeT5~\cite{Wang2021CodeT5IU}, SantaCoder~\cite{allal2023santacoder}, StarCoder~\cite{Li2023StarCoderMT} and CodeLlama~\cite{Rozire2023CodeLO}, adopt BPE as the tokenization method.
Further enhancements to BPE for source code have been proposed.
For example, Chirkova~\cite{Chirkova2023CodeBPEIS} suggests that clustering punctuation characters into single tokens can reduce average token length by 17\% without impacting model performance.
Notably, even though the tokenizers are optimized for source code, they still need to deal with the unnecessary tokens introduced by the human-centric grammar.
AI-oriented grammar optimizes the representation of source code in a more fundamental way, which is orthogonal to these existing tokenization methods.
\vspace{-1mm}
\section{Threats to Validity}
\label{sec:threat}
\noindent \textbf{Constrained Model Selection}
\su{Our experimental scope in RQ2 is restricted by our computational resources, limiting our evaluation to models with around 1B parameters.
These models are relatively modest in scale.}
However, while the model size is expanding, the fundamental issue of computation waste caused by human-centric code grammar remains unaddressed. 
Therefore, the insights derived from our experiments with smaller models are still highly relevant for understanding inefficiency issues in larger models.

\noindent \textbf{Limited Programming Language}
Our research primarily investigates the implementation of AI-oriented grammar in Python, a language widely utilized by existing LLMs for programming tasks. 
This initial exploration has shown that AI-oriented grammar effectively reduces computational costs during inference. 
However, \su{the conclusions drawn from Python may not generalize to other programming languages.}
We thus leave the exploration of its implementation in other languages as future work.

\noindent \textbf{Inefficient Implementation}
We implement a proof-of-concept converter to convert the code between \textsc{SimPy} and Python.
While this converter provides seamless translation, its efficiency is not optimized.
For instance, it is developed in Python, which is less efficient compared to languages like C++.
This aspect could potentially result in an underestimation of the converter's performance in our experimental evaluations.

\vspace{-1mm}
\section{Discussion}
\label{sec:discussion}

\noindent \textbf{Limitations in practice}
Though extending the applicability of AI-oriented grammar, DualCode relies on a rule-based converter.
The converter, we implemented for \textsc{SimPy}, is AST-based, which implicitly requires the input and output code of LLMs under the DualCode framework to satisfy the grammar correctness.
For the output, grammar correctness is a fundamental expectation for a qualified LLM-based assistant.
Thus, this requirement from DualCode is not an additional constraint set to the model but aligns with the goal of a reliable AI service.
However, it poses challenges when dealing with user-provided input, which may not always be grammatically correct.
It is not a concern to models handling natural-language-to-code tasks.
However, the requirement may limit the application of \textsc{SimPy} when some tasks involve partial source code as input, such as LLM-based code completion.
Addressing this limitation could involve developing an error-tolerant converter or grammar, which is a crucial direction for future research.

\noindent \textbf{Learning the AI-oriented grammar}
The learning of AI-oriented grammar could be a tricky task.
In our experiments, we demonstrate the effectiveness of fine-tuning AI models with \textsc{SimPy} using the next token prediction task.
However, this simple fine-tuning strategy requires a large number of \textsc{SimPy} samples, 100\% of the dataset in our experiments.
A more efficient adaptation process would significantly enhance the utility of AI-oriented grammar.
However, current research on how AI models learn code grammar is still limited.
Although studies~\cite{Wan2022WhatDT,Chen2022CATprobingAM, Ma2022IsSP} have shown that LLMs typically grasp code grammar knowledge in their initial layers, the exact learning mechanism remains unclear.
Therefore, a thorough analysis in this area is much needed.

\noindent \textbf{Utility of AI-oriented grammar}
\su{In this paper, we demonstrate the effectiveness of the sequential training scheme, where the model is initially trained with the original grammar and then the AI-oriented grammar.
It achieves an equivalent, or even improved, performance compared to the model trained merely with the original grammar.
Such a training method incurs an increase in the cost of the model training.
For example, training CodeGen on the original Python dataset costs 183,628 training steps, and 100,288 additional steps are taken during the further finetuning on the 100\% SimPy dataset.
Nevertheless, mastering AI-oriented grammar still reduces energy consumption in the long run. Training is performed only once or occasionally, while inference tasks can be continuous and massive after the system is deployed. The post-deployment operational cost is a primary component of the overall cost, sometimes reaching 90\% of total expenses~\cite{desislavov2023trends}. Consequently, despite the additional costs incurred during training, implementing AI-oriented grammar remains highly beneficial from a practical standpoint.
}

\vspace{-1mm}
\section{Conclusion and Future work}
In this paper, we, for the first time, propose the concept of AI-oriented grammar to address the inefficiency of AI coders in processing the code written in human-centric grammar.
Through an empirical study guided by three research questions, we successfully demonstrate the feasibility and potential of this novel concept.
During our research, we have developed the first-ever AI-oriented Python grammar.
Additionally, we introduced an inference framework designed to empower models to effectively process both AI-oriented and human-centric grammars. 

As an emerging field, AI-oriented grammar presents numerous unexplored questions.
For example, an interesting finding from our experiments is that models trained with AI-oriented grammar can even improve the model's accuracy in code generation tasks.
This emphasizes the critical role of grammar as a foundational element for LLMs in grasping code semantics.
Designing grammars that are inherently more comprehensible to AI models could significantly enhance their performance.
Our current research provides a preliminary insight into this aspect, opening doors for in-depth future studies.
Additionally, the process of simplifying grammar, as exemplified by our manual creation of \textsc{SimPy}, raises the question of whether an automated approach could create optimal grammar rules for AI models.
\su{A potential solution for simplifying the grammar could be iteratively searching for grammar tokens/structures that can be removed with the help of a parser generator.
Moreover, saving the training cost for teaching LLMs AI-oriented grammar is also of great practical value, where a more efficient training method for LLMs to learn new programming grammar is urgently needed.}
We, therefore, call for the software engineering community to engage further with this promising topic, recognizing its potential to revolutionize the field of AI coders.

\begin{acks}
We thank Dr Zhe Hou for the helpful discussion when preparing the manuscript.
This research / project is supported by Xiaoning Du’s Google Research Scholar Program Award and the National Research Foundation, under its Investigatorship Grant (NRF-NRFI08-2022-0002).
Any opinions, findings and conclusions or recommendations expressed in this material are those of the author(s) and do not reflect the views of National Research Foundation, Singapore.
\end{acks}

\bibliographystyle{ACM-Reference-Format}
\bibliography{sample}

\end{document}